\documentclass[aps,pra,twocolumn,amsmath,amssymb,floats,floatfix,showpacs]{revtex4-1}

\usepackage{amsmath}              
\usepackage{graphicx}
\usepackage{dcolumn}

\begin{document}

\title{Environment-dependent dissipation in quantum Brownian motion}   
\author{J. Paavola}
\affiliation{
Department of Physics and Astronomy, University of Turku,
FI-20014 Turun yliopisto, Finland
}
\author{J. Piilo}
\affiliation{
Department of Physics and Astronomy, University of Turku,
FI-20014 Turun yliopisto, Finland
}
\author{K.-A. Suominen}
\affiliation{
Department of Physics and Astronomy, University of Turku,
FI-20014 Turun yliopisto, Finland
}

\author{S. Maniscalco}
\affiliation{
Department of Physics and Astronomy, University of Turku,
FI-20014 Turun yliopisto, Finland
}

\date{\today}    

\begin{abstract}
The dissipative dynamics of a quantum Brownian particle is studied for different types of environment. We derive analytic results for the time evolution of the mean energy of the system for Ohmic, sub-Ohmic and super-Ohmic environments, without performing the Markovian approximation. Our results allow to establish a direct link between the form of the environmental spectrum and the thermalization dynamics. This in turn leads to a natural explanation of the microscopic physical processes ruling the system time evolution both in the short-time non-Markovian region and in the long-time Markovian one. Our comparative study of thermalization for different environments sheds light on the physical contexts in which non-Markovian dissipation effects are dominant.

\end{abstract}

\maketitle

\section{Introduction}\label{intro}

The study of open quantum systems has recently received renewed attention due to the importance of environment-induced effects, such as quantum decoherence, both in fundamentals of quantum theory and in newly emerging quantum technologies \cite{MEderived,weiss}.
Quantum systems are never completely isolated from their external environment. The interaction between the system and its surroundings induces decoherence phenomena destroying quantum superposition and
entanglement. Environment induced decoherence has been studied
extensively for the damped harmonic oscillator model. In this context, the decoherence of a superposition of Gaussian wave packets, a prototype of
Schr\"odinger cat state, has been studied theoretically and experimentally  \cite{EID,Scats,zurekdecoherence,engineeredres,Haroche96a}.

Decoherence also plays a major role in quantum information technologies since
the coherence time typically determines the operational time of a quantum device, e.g., a quantum logic gate. Understanding the dynamics of exemplary quantum systems interacting with their environment is therefore of crucial importance for both fundamentals and applicative aspects of quantum theory.

In this paper we investigate the dynamics of a damped harmonic oscillator, or harmonic quantum Brownian motion (QBM) model, interacting with different thermal bosonic environments.
The QBM model is one of the few models of open quantum systems
amenable to an analytical solution \cite{exactVSourME,weiss,MEderived,Feynman63a,Caldeira83a,Haake85a,Grabert88a,numericalcomparison,analytic_solution,rwa,lindblad,einselection,Eisert,Paz93,Maniscalco04a,JOPBManiscalco,Haake96,paper1,paper3}.

A very general derivation of the nonlinear Langevin equations for a damped harmonic oscillator, for general microscopic system-environment couplings, is given in Ref. \cite{paper11}. In the case of bilinear system-environment coupling the generalized exact master equation for the reduced system is known as the Hu-Paz-Zhang master equation \cite{numericalcomparison}. This master equation is typically solved numerically. In some cases, explicit solutions in closed form exist, e.g., for an initial Gaussian wave packet or a superposition of Gaussian wave packets  \cite{paper10}. In this paper we study the dynamics for initial Fock states of the harmonic oscillator and we use a perturbative approach that nonetheless allows to study non-Markovian features due to structured environments. Our aim is indeed to obtain simple analytical expressions in closed form for the observables of interest in order to gain insight in the fundamental microscopic processes ruling the non-Markovian dissipative dynamics.

The QBM model is widely used in many physical contexts. Indeed it describes a quantum
electromagnetic field propagating in a linear dielectric medium
\cite{applicationogQBM}, a particle interacting with a quantum field in dipole
approximation \cite{einselection} and a single trapped ion subjected to artificial colored noise \cite{Maniscalco04a}. In
addition to these quantum optical applications, the QBM model is
used in nuclear physics \cite{nuclear} and quantum chemistry
\cite{chemistry}.
For this reason the literature on this model is vast and crosses several fields of science. Interesting results have recently led to a better understanding of anomalous diffusion for the free QBM model, i.e., in absence of a trapping potential \cite{paper6}. The dynamics of an initial Gaussian state in an anharmonic potential have also been studied \cite{paper2}.

The Hu-Paz-Zhang master equation has recently been used  to study the entanglement dynamics of initial coherent and twin-beam states of two non-interacting harmonic oscillators linearly coupled to common \cite{paper4,paper4a,paper4b} or independent structured reservoirs \cite{paper9a,paper9}. For independent reservoirs the entanglement dynamics for initial non-Gaussian states was presented in \cite{paper7}. The non-Markovian dynamics of entanglement for two coupled harmonic oscillators was investigated in  \cite{paper5} for different types of environment. Very recently strategies of optimal decoherence control have been demonstrated for non-Markovian two-level systems  \cite{paper8}. We focus in this paper on a system more complicated than a two-state system, i.e., a single quantum harmonic oscillator, but it would certainly be of interest to extend the optimal control analysis to systems with a non-finite Hilbert space.


In this paper we use a time-convolutionless perturbative master equation that does not rely on the Markovian approximation and can therefore describe situations in which the spectrum of the environment has a structure. This is, e.g., the case of atom lasers \cite{atomlaser} or atoms decaying in
photonic band gap materials \cite{useNONmarkovian}. By specifying the form of the spectrum we obtain analytical expressions for the mean energy of the system in a close form. In this way we can establish a clear connection between the reservoir spectral properties and the non-Markovian dynamics of the quantum Brownian particle.

The  coupling between the system and the quantized oscillators constituting the bosonic reservoir is
given, in the continuum limit, by the reservoir spectral density. Different physical contexts are characterized by different forms of the reservoir spectrum. The three main classes typically considered in the literature are the so-called Ohmic, sub-Ohmic and super-Ohmic spectra.
We will compare the thermalization process for these three types of reservoirs.

It is worth mentioning that recent advances in reservoir engineering techniques \cite{engineeredres} pave the way to experiments aimed at simulating paradigmatic models of open quantum systems as the one considered in this paper. In the trapped ion context, e.g., the simulation of a QBM model for an Ohmic environment has been proposed in Refs. \cite{Maniscalco04a,simulator}. The same method can be extended straightforwardly to simulate the sub-Ohmic and super-Ohmic environments here considered. These experiments would allow to test in a controlled way a fundamental and ubiquitous model such as QBM. Understanding which type of environment leads to the faster or slower decoherence/dissipation dynamics can be of great importance in the choice of the physical system for implementing realistic quantum devices such as a quantum computer.

%


%

%


The paper is structured in the following way. In Sec. \ref{ME} we introduce
the model under study and the master equation describing
the dynamics. Section \ref{environment} introduces the three different
examplary reservoirs, namely the Ohmic, sub-Ohmic and super-Ohmic reservoirs, used in our comparative study. In Secs. \ref{decayrates} and \ref{heating} we discuss our results for the decay rates
and heating dynamics of the QBM model. Finally, in Sec. \ref{final} we present
conclusions and outline possible future prospects.
\section{Master equation for QBM}\label{ME}
Let us consider a quantum particle of mass
$m$ moving in a harmonic potential. The Hamiltonian of
the system is
\begin{equation}\label{eq:systemhamilton}
H_S=\omega_0\left(a^\dagger a+\frac{1}{2}\right),
\end{equation}
where $a$ and $a^\dagger$ are the creation and annihilation operators of the quantum harmonic oscillator, $\omega_0$ is the frequency and $\hbar$ is set to $1$. The environment is a heat bath modeled as an infinite chain of
harmonic oscillators
\begin{equation}\label{eq:reservoirhamiltonian}
H_E=\sum_{n=0}^{\infty}\omega_n\left(b_n^\dagger b_n
+\frac{1}{2}\right),
\end{equation}
where $b_n$ and $b_n^\dagger$ are the creation and annihilation operators, respectively, and $\omega_n$ is the frequency of the $n$th
oscillator. The system and the reservoir are coupled
linearly via the position operators, $X\propto a+a^\dagger$ and $x_n\propto b+b^\dagger$ for the
system and reservoir oscillators, respectively, so that the
interaction Hamiltonian is given by
\begin{equation}\label{eq:interactionhamilton}
H_I=
\frac{1}{\sqrt{2}}(a+a^\dagger)\sum_n k_n
(b_n+b_n^\dagger),
\end{equation}
where $k_n$ measures the coupling between each reservoir mode and the
system oscillator.

A master equation describing the QBM dynamics can be derived starting from the total Hamiltonian
\begin{equation}
H=H_S+H_E+\alpha H_I,
\end{equation}
where $\alpha$ is a dimensionless constant proportional
to the strenght of the coupling between the system and the
environment. In the weak couling limit (i.e., when $\alpha\ll1$),
assuming initially factorized state ($\rho=\rho_S\otimes\rho_E$) and
a thermal reservoir, we obtain the following secularly approximated
master equation for the damped harmonic oscillator \cite{MEderived}
\begin{align}\label{eq:mainME}
\frac{d}{dt}\rho_S(t)=&\frac{\Delta(t)-\gamma(t)}{2}\left(2a^\dagger
\rho_S\nonumber a-a a^\dagger\rho_S-\rho_S a
a^\dagger\right)\\
&+\frac{\Delta(t)+\gamma(t)}{2}\left(2a\rho_Sa^\dagger-a^\dagger
a\rho_S-\rho_Sa^\dagger a\right),
\end{align}
where
\begin{align}
\label{eq:delta}
\Delta(t) =& 2\int_0^{t}dt'\,
\int_0^{\infty}d\omega\,J(\omega)\left[N(\omega)+\frac{1}{2}\right]\\
&\times\cos(\omega\nonumber
t')\cos(\omega_0 t'), \\
\gamma(t) =& 2\int_0^{t}dt'\,
\int_0^{\infty}d\omega\,\frac{J(\omega)}{2}\sin(\omega
t')\sin(\omega_0 t').\label{eq:pikkugamma}
\end{align}
In the equation above $N(\omega) =
(e^{ \omega/k_B T}-1)^{-1}$ is  the average number of reservoir thermal excitations , with $k_B$  the Boltzmann  constant and $T$ the
reservoir temperature, and $J(\omega)$ is the
spectral density of the environment defined, in the continuum limit, as
\begin{equation}
J(\omega)=\alpha^2\sum_n\frac{k_n^2}{m_n\omega_n}\delta(\omega-\omega_n),
\end{equation}
with $m_n$ the masses of the environmental oscillators.
In
deriving the master equation no Markovian approximation has been
done. The memory effects are included in the time-dependent
coefficients $\Delta(t)$ and $\gamma(t)$. The latter term is known as dissipation coefficient and gives rise to a
classical damping term that is not dependent on temperature. The former term $\Delta(t)$ is known as diffusion coefficient and is directly proportional to the reservoir temperature \cite{MEderived}.

It is worth mentioning here that performing a secular approximation does not affect the non-Markovian short time dynamics of certain observables in the weak coupling limit \cite{analytic_solution}.
In this paper we focus on the dynamics of one of such observables, namely the heating function.

\section{Modeling different types of reservoirs}\label{environment}
We now introduce a class of spectral densities in order to compare the QBM dynamics for different  types of reservoirs. The spectral densities we examine are of the form
\begin{equation}\label{spectra}
J(\omega)=\alpha^2\omega_{c}^{1-s}\omega^s e^{-\omega/\omega_c}.
\end{equation}
The exponential cutoff is introduced to eliminate divergencies in the $\omega \rightarrow \infty$ limit. We have compared various types of cutoff functions and concluded that their different analytical forms do not play a major role in the dynamics of QBM.
For the sake of simplicity we therefore focus, in the rest of the paper, on the exponential cutoff. The parameter $s$ appearing in Eq. (\ref{spectra}) is a constant that can acquire values $<1$, $1$ or $>1$, corresponding to the so called sub-Ohmic, Ohmic and super-Ohmic spectral densities, respectively. In this paper we consider some examples and fix the value of $s$ to $1/2$, $1$ and $3$. The three cases have different physical interpretations.
The Ohmic spectrum ($s=1$) gives, for QBM, a friction-like force
that is proportional to velocity. The Ohmic spectrum can be used,
e.g., to describe charged interstitials (conductive electrons) in
metals \cite{weiss}. The super-Ohmic spectral density ($s=3$) corresponds to, e.g., a phonon bath in
one or three dimensions, depending on the symmetry properties of the
strain field \cite{weiss}. It is also possible to show that this
type of environment can be used in describing the effect of the
interaction between a charged particle and its own electromagnetic
field \cite{superohmic}. The sub-Ohmic spectral density ($s=1/2$ ) corresponds to the type of noise that may occur in some solid state devices and, in the
high $T$ case, is similar to the "1/f noise" in Josephson junctions
\cite{similarto1/f}.

We introduce the spectral distribution given by
\begin{equation}\label{eq:spectraldistr}
I(\omega)=J(\omega)\left[N(\omega)+\frac{1}{2}\right].
\end{equation}
This quantity contains all the information needed, in the weak coupling limit, about the reservoir, i.e.,  the density of modes and the occupancy of each mode.  The
spectral distribution depends on the temperature of the reservoir
through the average number of reservoir thermal excitations $N(\omega)$.
%
At high temperatures $T$ we can approximate $N(\omega) \approx k_B T/ \omega $
%
%
while at zero temperature $N(\omega)=0$. In the rest of the paper we discuss the QBM dynamics in these two temperature regimes.

\begin{figure}
\includegraphics[width=8cm]{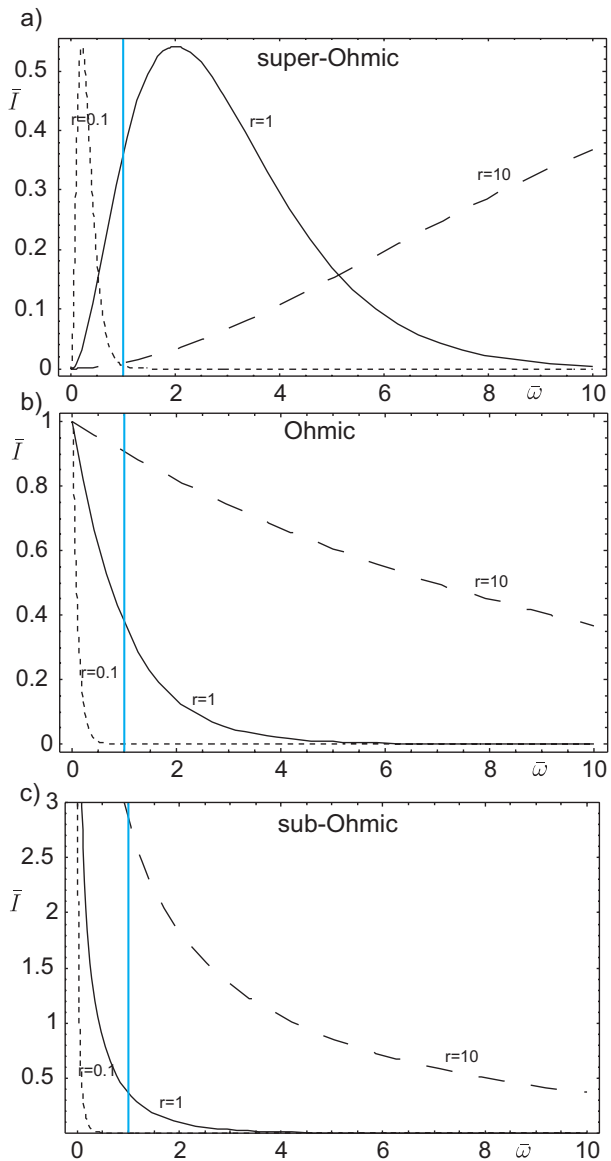}
\caption{\label{tab:spectra}Spectral distributions of different
reservoirs at high temperatures. Here $\bar{I}=I/(\alpha^2 k_BT)$
and $\bar{\omega}=\omega/\omega_0$. For each spectral curve the
location of the cutoff frequency is given by $\bar{\omega}_c=\omega_c/\omega_0 = r$. The location of the oscillator frequency has been marked with a
solid vertical line.}
\end{figure}

\begin{figure}
\includegraphics[width=8cm]{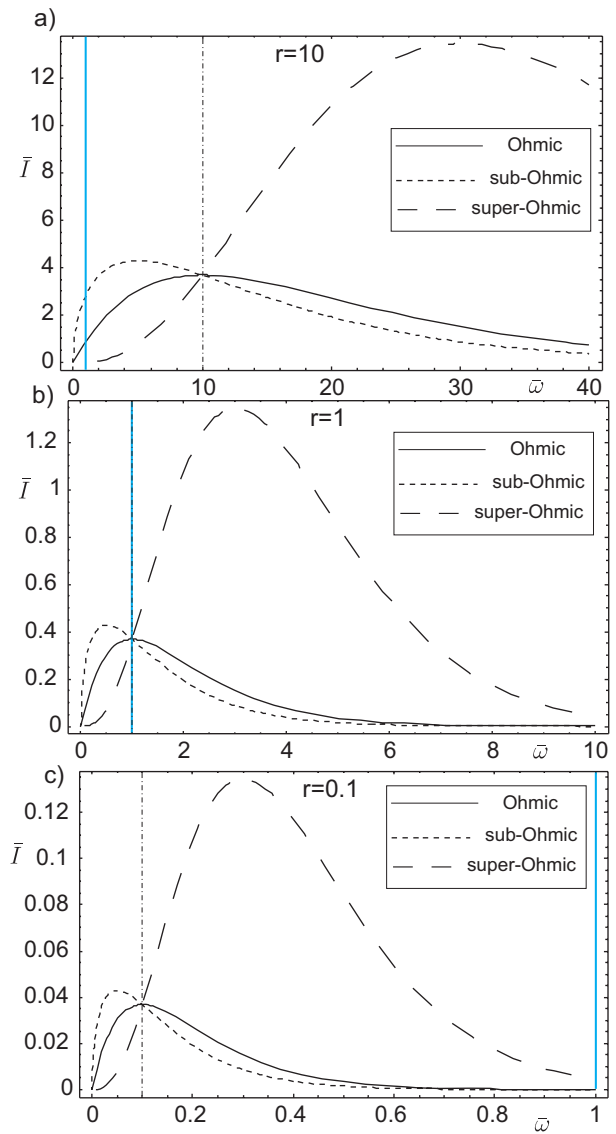}
\caption{\label{fig:zerospectrum}Spectral distributions at zero
temperature grouped according to the resonance parameter $r$. Here
$\bar{I}=2I/(\alpha^2\omega_0)$ and $\bar{\omega}=\omega/\omega_0$. The vertical dotted-dashed line is the location of the cutoff frequency $\bar{\omega}_c=\omega_c/\omega_0$.}
\end{figure}

A relevant parameter in the description of QBM is the resonance parameter $r$ defined as the ratio between the cutoff frequency $\omega_c$ and the frequency of the system oscillator $\omega_0$, i.e.,
\begin{equation}\label{eq:r}
r=\frac{\omega_c}{\omega_0}.
\end{equation}
We consider three exemplary values of the resonance parameter,
namely $r$ equal to $0.1$, $1$ and $10$. The $r=0.1$ case is
characterized by the fact that the effective coupling between the system oscillator
and the environment is very small for all three reservoir types because the system oscillator is detuned from the peak of the reservoir spectral distribution. We call this the off-resonant
case. The Ohmic and sub-Ohmic reservoirs are such that the effective coupling
between the system and the reservoir becomes stronger when $r$ grows from $0.1$ to
$10$. The super-Ohmic reservoir, on the contrary, shows the highest effective coupling for $r=1$, while the  $r=0.1$ and  $r=10$ cases correspond to relatively weak couplings. Plots of the spectral distribution for different values of $r$ in the high and zero temperature limits are shown in Figs.  \ref{tab:spectra} and  \ref{fig:zerospectrum}.

The spectral distribution at high temperatures, given by
\begin{equation}\label{eq:IhighT}
I(\omega)=\alpha^2k_BT\left(\frac{\omega}{\omega_c}\right)^{s-1}e^{-\omega/\omega_c},
\end{equation}
%
%
is shown in Fig. \ref{tab:spectra}.
The sub-Ohmic spectrum has a divergency point
at $\omega=0$. This causes large effective coupling induced by the
low frequency part of the spectral density. An opposite example can
be found for the super-Ohmic reservoir, where the peaks of the spectrum
in the cases $r=1$ and $r=10$, lie in the higher frequency range.


The spectral density for zero $T$ reservoirs, given by
\begin{equation}\label{eq:IzeroT}
I(\omega)=\frac{\alpha^2}{2}\omega_{c}^{1-s}\omega^se^{-\omega/\omega_c},
\end{equation}
is shown in Fig. \ref{fig:zerospectrum} (the plots are here grouped according to parameter
$r$ for clarity).
%
A key difference with respect to the high temperature case is that at zero
temperature, the sub-Ohmic spectrum does not diverge in zero anymore.

As we will see in Secs. IV and V, the QBM heating dynamics depends crucially on the different form of the spectral distributions in the Ohmic, sub-Ohmic and super-Ohmic cases. Our main goal is to establish a clear connection between the reservoir properties and the dynamics of both the decay rate and the heating function, as done, for an Ohmic reservoir with Lorentzian cutoff, in Ref.  \cite{simulator}. In this way we will be able to motivate from a physical point of view the origin of the different QBM dynamics for different reservoirs, and therefore in different physical contexts.

\section{Decay rates}\label{decayrates}
The frontfactors $[\Delta(t)+\gamma(t)]/2$ and
$[\Delta(t)-\gamma(t)]/2$ in the master equation \eqref{eq:mainME} represent the
relaxation rates for the two decay channels of the QBM model. In the
Fock state basis the former rate is associated to $|n\rangle\rightarrow|n-1\rangle$ transitions, i.e. to the transfer of one excitation from the system to the environment. The latter rate corresponds to $|n\rangle\rightarrow|n+1\rangle$ transitions, i.e. it describes the absorption of one excitation from the environment.
These transitions,
describing heating or cooling of the quantum harmonic
oscillator due to the interaction with the external environment, destroy the quantum coherence of initial superpositions.

After a certain reservoir-dependent time, the decay rates reach their constant positive Markovian values
\begin{eqnarray}
\label{eq:delta}
\Delta_M &=&  \pi I(\omega_0), \\
\label{eq:sab3} \gamma_M &=& \frac{\pi}{2}J(\omega_0).
\end{eqnarray}
The decay rates
can also temporarily attain negative values. When this happens, the
corresponding decay channel has been shown to operate in a reverse
way, i.e., the down channel actually induces heating and vice versa
\cite{negativejumps}. We will now examine the decay rates in
non-Markovian time scales at high and zero temperatures.
\subsection{High temperatures}

%
\begin{figure*}
\includegraphics[width=8cm]{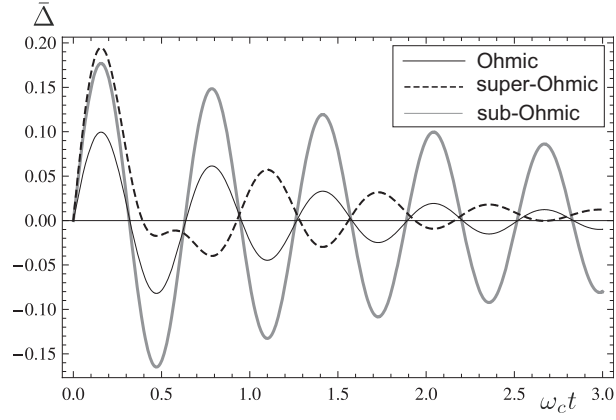}
\caption{\label{fig:example_decayrate}Decay rates at high
temperatures for different types of reservoir in the non-Markovian
time scales. Here $\bar\Delta=\Delta/(2\alpha^2k_BT)$ and
$r=0.1$}
\end{figure*}

\begin{figure*}
\includegraphics[width=15cm]{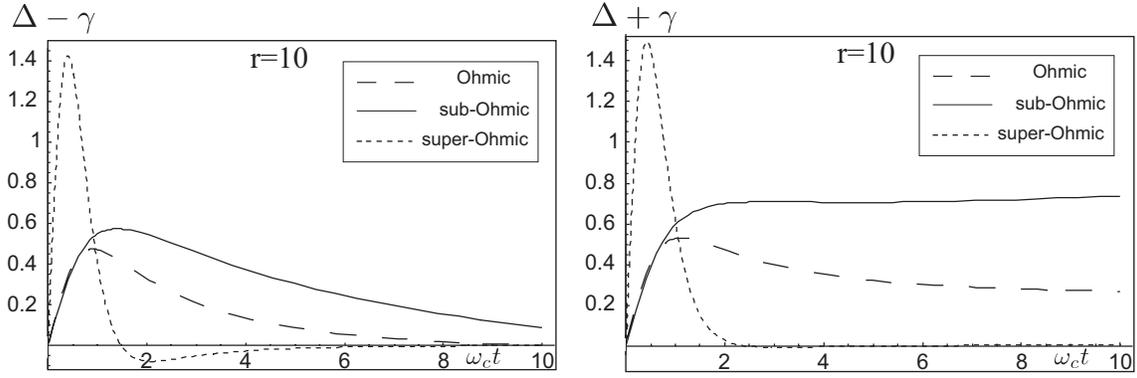}
\caption{\label{fig:channels10}The decay rates at zero temperature
for  $r=10$. The figure on the left depicts the
transition up rates, while the figure on the right corresponds to the
transition down rates.}
\end{figure*}

%
At high temperatures $\Delta(t)\gg\gamma(t)$, and for
time scales much shorter than the thermalization time, both the transition up and
down channels operate at approximately the same rate $\Delta(t)/2$. We have obtained an analytic expression for  $\bar{\Delta}(t)=\Delta(t)/(\alpha k_BT)$ for all the three types of environment considered in this paper. More precisely, for the Ohmic environment we obtain
\begin{align}
\bar{\Delta}(t)=-i &\cosh\left(\frac{1}{r}\right)\left[ ci\left(\frac{z}{r}\right)-ci\left(\frac{z+2t}{r}\right)\right]\\
&+\sinh\left(\frac{1}{r}\right)\left[ si\left(\frac{z}{r}\right)-si\left(\frac{z+2t}{r}\right)\right],\nonumber
\end{align}
and for the super-Ohmic environment
\begin{align}
\bar{\Delta}(t)=&\frac{4t\cos\left(\frac{t}{r}\right)}{(1+t^2)^2}-\frac{2\sin\left(\frac{t}{r}\right)}{r+rt^2}+\frac{1}{r^2}\\
&\times\bigg\{-i\cosh\left(\frac{t}{r}\right)\left[ci\left(\frac{z}{r}\right)\nonumber
-ci\left(\frac{z+2t}{r}\right)
\right]\\
&+\sinh\left(\frac{t}{r}\right)\left[si\left(\frac{z}{r}\right)-si\left(\frac{z+2t}{r}\right)\right]\bigg\},\nonumber
\end{align}
where $z=i-t$, $ci(x)$ and $si(x)$ are the cosine and sine integrals defined as
$
ci(x)=-\int_x^{\infty} \frac{\cos(x)}{x}dx$ and $si(x)=-\int_x^{\infty} \frac{\sin(x)}{x}dx$.
Finally, the decay rate for the sub-Ohmic environment is
\begin{align}
\bar{\Delta}(t)&=-\frac{2\pi e^{-1/r}}{\sqrt{2i-2t}(1+t^2)^{1/4}}
\left(\frac{1}{4}+\frac{i}{4}\right)\nonumber
\sqrt{\frac{r(1+it)}{\sqrt{1+t^2}}}\\
&\times
\bigg\{
\sqrt{1+t^2}\,\mathrm{erf}(z^-)-ie^{2/r}\sqrt{1+t^2}\,\mathrm{erf}(iz^-)\\
&+i\sqrt{z}\sqrt{z+2t}
\left[\mathrm{erf}(z^+)-ie^{2/r}\mathrm{erf}(iz^+)
\right]\nonumber
\bigg\},
\end{align}
where $z^\pm=[(1+i)\sqrt{i\pm t}]/\sqrt{2r}$ and $\mathrm{erf}(x)=\frac{2}{\sqrt{\pi}}\int_0^z e^{-t^2}dt$.

From the analytic expressions of the diffusion coefficient $\bar{\Delta}(t)$ one can show that such a quantity oscillates taking temporarily negative values for $r \ll 1$ for all reservoir types. The origin of these oscillations can be traced back to the form of the spectral distribution. Figure 1 shows that, e.g., for $r=0.1$ most part of the spectrum is located in a region of the frequency space such that $\omega < \omega_0$. Figure 3 shows the oscillatory behavior of $\bar{\Delta}(t)$ for this value of $r$. As $r$ grows, the spectrum starts to overlap with $\omega_0$. We note that, in the case of the sub-Ohmic environment, the diffusion coefficient presents oscillations for all values of $r$, but it attains negative values only when $r\ll1$.

The Markovian value of the diffusion coefficient $\Delta(t)$ is proportional to $I(\omega_0)$ [See Eq.(\ref{eq:delta})]. As one can see from Fig. \ref{tab:spectra}, for  $r=10$ the highest value of $\Delta_M$ corresponds to the sub-Ohmic environment while in the off-resonant $r=0.1$ case $\Delta_M$ is small for all reservoir types.

\subsection{Zero temperature}\label{decayzeroT}
When the system oscillator interacts with a zero-temperature reservoir, the Markovian theory predicts that energy is transferred from the system to the environment, i.e., the oscillator is driven towards its ground state. Our non-Markovian theory, however, shows that for times $\omega_c t \ll 1$, the average energy of the system oscillator may increase, as we will see in detail in Sec. V. This is due to the form of the interaction Hamiltonian, given by Eq.
\eqref{eq:interactionhamilton},  containing
four terms characterizing the emission and absorption processes,
namely, $ab_n$, $ab_n^\dagger $, $a^\dagger b_n$ and $a^\dagger
b_n^\dagger$. The two terms in the middle correspond to real processes conserving the unperturbed energy, while the other two are known as the counter rotating terms. These terms describe the simultaneous creation or annihilation of a quantum of energy both in the system and in the reservoir oscillators. The energy required for such processes to occur comes from the system-reservoir coupling. By combining
these two counter rotating terms, we obtain a process that
corresponds to an energy conserving process. It has been shown that at zero temperature the dynamics of the decay rate for the transitions up originates from
these counter rotating terms \cite{rwa}.

The decay rates at zero temperature show similar dependence on the parameter $r$ as the one discussed in the high temperature case. In particular, for $r\ll1$ both $\Delta(t)+\gamma(t)$ and $\Delta(t)-\gamma(t)$ oscillate attaining negative values, a clear signature of the non-Markovian  behavior of the system. Compared to the high $T$ case, oscillations in $\Delta(t)-\gamma(t)$ obtaining negative values are present also for higher values of $r$, e.g. $r=1$, for all reservoir types. For the super-Ohmic environment the decay rates show a strong initial jolt for all values of $r$. Having in mind the form of the spectral distributions one sees that the initial jolt is present whenever the peak of the spectrum lies in the frequency region $\omega > \omega_c$.

For  $\omega_c t \gg 1$ the decay rate $\Delta(t)-\gamma(t)$, describing $\vert n \rangle \rightarrow \vert n+1 \rangle$ transitions in the system oscillator, approaches zero as expected from the Markovian theory, while $\Delta(t)+\gamma(t)$, describing $\vert n+1 \rangle \rightarrow \vert n \rangle$ transitions, reaches a constant positive value $\Delta_M+\gamma_M=\pi I(\omega_0)$, as shown in Fig. \ref{fig:channels10} for the $r=10$ case.

An interesting aspect worth mentioning is visible in the super-Ohmic case for $r=10$. Figure  \ref{fig:channels10} clearly shows that for the super-Ohmic spectrum the
decay rate $\Delta(t)+\gamma(t)$, i.e. the cooling rate of the system oscillator, approaches its small but nonzero constant value already for $\omega_c t \approx3$, while at the same time the decay rate $\Delta(t)-\gamma(t)$, describing heating of the system oscillator, attains negative value and tends to zero while remaining negative.

It has been shown in Ref. \cite{negativejumps} that in correspondence to
negative regions of the time dependent coefficients, in our case $\Delta(t)\pm\gamma(t)$, reverse transitions
restoring the previous quantum state occur. In view of these
results one can argue that, in the case considered above, the up channel acts like a transition down channel.
This implies that the thermalization is achieved via a reverted
transition up channel, while the actual transition down channel is
almost completely closed.

In the next section we will see how the behavior of the decay rates is related to the heating dynamics of the quantum Brownian particle and we will investigate the differences in the dynamics of the mean energy of the system due to different environments.

\section{Heating of a quantum Brownian particle}\label{heating}

\subsection{Markovian thermalization dynamics}
The QBM dissipative dynamics can be described by means of the heating
function, defined as
\begin{equation}
\langle n\rangle=a^\dagger a.
\end{equation}
The analytical expression for the heating function is given by
\cite{heatingfunctionderived,lindblad}
\begin{equation}\label{eq:heatingfunction}
\langle n(t)\rangle=e^{-\Gamma(t)}\langle
n(0)\rangle+\frac{1}{2}\left[e^{-\Gamma(t)}-1\right]+\Delta_\Gamma(t),
\end{equation}
where $\Gamma(t)$ and $\Delta_\Gamma(t)$ are defined as
\begin{eqnarray}
\label{eq:sab1}  \Gamma(t)&=&2\int_0^t \gamma(t_1)dt_1\\
\label{eq:sab2} \Delta_\Gamma(t)&=&e^{-\Gamma(t)}\int_0^{t}e^{\Gamma(t_1)}\Delta(t_1)dt_1.
\end{eqnarray}
%

\begin{figure}
\includegraphics[width=8cm]{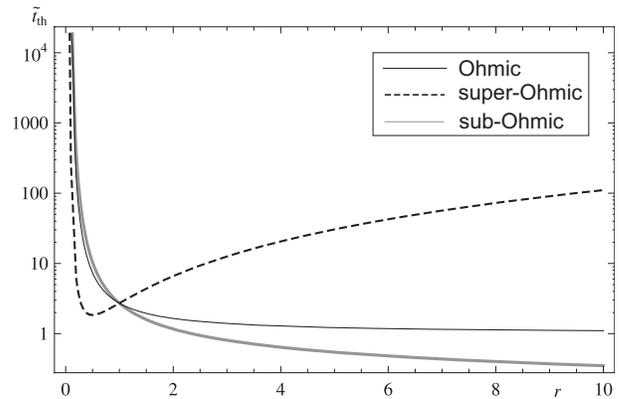}
\caption{\label{fig:thermalization}Thermalization times for different reservoirs. Here $\tilde{t}_{th}=\pi\alpha^2 t_{th}$.}
\end{figure}

In the following we focus on the case where the oscillator is initially in the
ground state, i.e., $\langle n(0)\rangle=0$, and perform a comparative study of the heating function dynamics for different reservoir structures.
%


We begin by looking at the Markovian dynamics describing the time evolution for times much greater than the reservoir correlation time. The Markovian expression of the heating function $ \langle n(t)\rangle_M$ is obtained inserting the Markovian expressions of the diffusion and dissipation coefficients, given by Eqs. \eqref{eq:delta}-\eqref{eq:sab3}, into Eqs. \eqref{eq:heatingfunction}-\eqref{eq:sab2},
\begin{equation}\label{eq:longtimen}
\langle n(t)\rangle_M=N(\omega_0)\left(1-e^{-\Gamma t}\right),
\end{equation}
where $\Gamma=2\gamma_M= \pi J(\omega_0)$. From Eqs. \eqref{spectra} and \eqref{eq:sab3} we can express the reservoir thermalization time, in units of $\omega_0$, as follows
\begin{equation}\label{eq:therm}
t_{th} = \omega_0 / \Gamma = (\pi \alpha^2 )^{-1} r^{s-1} e^{1/r} .
\end{equation}

From Fig. \ref{fig:thermalization} we see that, for both the sub-Ohmic and the Ohmic reservoirs, the  thermalization time decreases monotonically for increasing values of $r$, i.e. for increasing values of the cutoff frequency with  respect to the frequency of the system oscillator $\omega_0$. On the contrary, for the super-Ohmic reservoir, there exist a value of $r$, namely $r\approx0.5$, minimizing the thermalization time. In general, all the three reservoir types considered in this paper are such that the thermalization time grows rapidly when $r\rightarrow 0$, and correspondingly the thermalization process is notably slowed down.
Our analysis suggests that, by appropriately changing the cutoff frequency of a high temperature engineered reservoir, it is possible to control the thermalization dynamics.

\subsection{Non-Markovian heating}

We now look at the non-Markovian short time dynamics of the heating function.
For times much smaller than the thermalization time Eq.
\eqref{eq:heatingfunction} can be approximated by
\begin{equation}\label{eq:appron}
\langle
n(t)\rangle=\int_0^t\left[\Delta(t_1)-\gamma(t_1)\right]\,dt_1.
\end{equation}
This equation establishes a clear connection between the heating function dynamics and the time dependent decay rate $\Delta(t)-\gamma(t)$ corresponding to transitions increasing the system oscillator energy, i.e., absorption of quanta from the environment. This is clearly related to our choice of the initial
condition $\langle n(0)\rangle=0$. In this case indeed, for times much shorter than the thermalization time, the absorption of a quantum of energy from the environment (heating) dominates over the opposite process, i.e., the emission of a quantum of energy into the environment (cooling).

\subsubsection{High Temperatures}

For high
temperature reservoirs, Eq. \eqref{eq:appron} can be further
approximated by \cite{heatingion}
\begin{equation}\label{eq:highTn}
\langle n(t)\rangle\approx\int_0^t\Delta(t_1)\,dt_1.
\end{equation}
The sign
of $\Delta(t)$ determines whether the heating
function grows monotonically or exhibits an oscillatory behavior. In more detail, when $\Delta(t)$ oscillates taking negative values then the heating function oscillates.

The oscillations in the heating function appear when the system
gives back to the reservoir some of the energy that had previously been absorbed from it. In other words, the direction of the energy flow
is reversed during the time periods in which the slope of $\langle
n(t)\rangle$ is negative. These oscillations are a sign of the
non-Markovian dynamics and are due to the finite reservoir memory that allows the
partial and temporary recovery of some of the information/energy lost in the
reservoir.

Equation (\ref{eq:highTn}) links the heating function to the diffusion coefficient dynamics. In the previous section we have seen how the structure of the reservoir, and in particular some specific system-reservoir parameters, determine the temporal behavior of $\Delta(t)$. By means of Eq. (\ref{eq:highTn}) we can now establish a connection between the reservoir spectrum and the heating process and compare the non-Markovian heating for the sub-Ohmic, Ohmic and super-Ohmic reservoirs.

\begin{figure*}
\includegraphics[width=15cm]{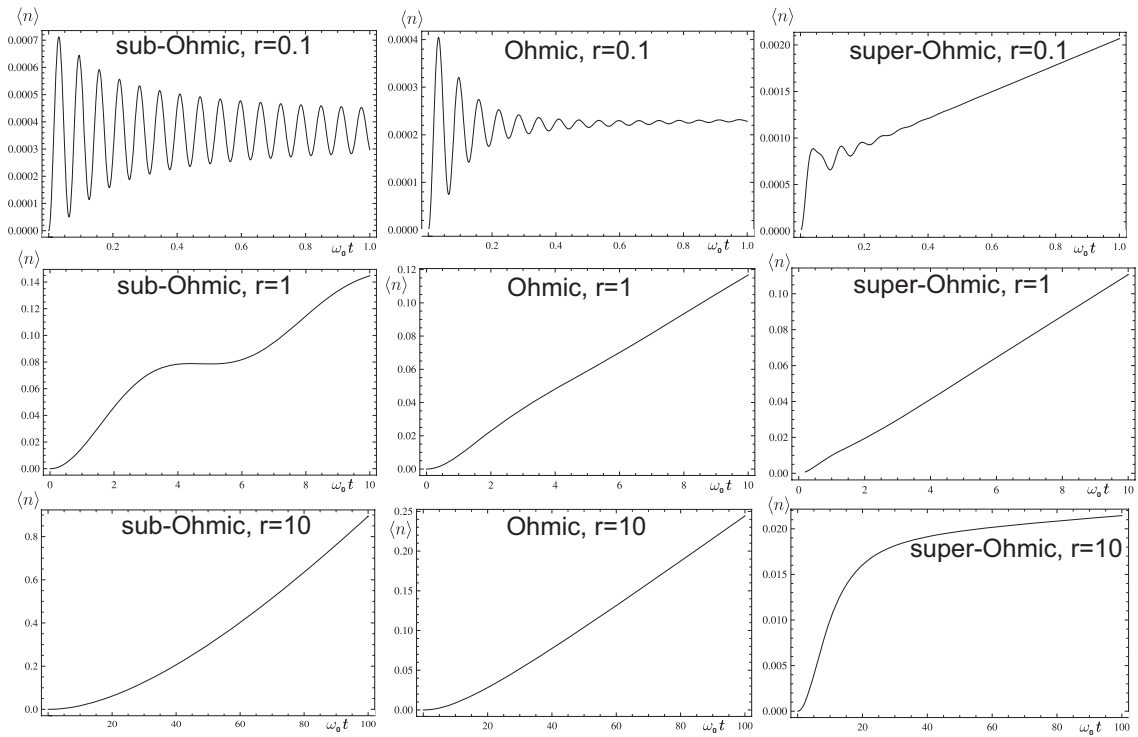}
\caption{\label{fig:nshort}Short time dynamics of the heating
function for different high temperature
reservoirs. In the plots $\alpha=0.01$ and we have set $\frac{k_BT}{\hbar\omega_0}=100$.}
\end{figure*}

From Fig. 6 one sees clearly that the non-Markovian heating, described by Eq. (\ref{eq:highTn}), presents two main types of behavior: an oscillatory behavior and a monotonic growth similar to the Markovian heating.
In Ref. \cite{simulator} it was demonstrated that, for an Ohmic reservoir and for $r=0.1$, oscillations in $\langle n(t)\rangle$ originate from the low frequency part of the spectrum while the monotonic heating is caused by the resonant part of the spectrum, namely by the  value of $I(\omega_0)$.

This connection between the features of $\langle n(t)\rangle$ and the spectrum seems to hold also for the three reservoir types discussed in this paper. Indeed, for $r\ge 1$ the decay rates $\Delta(t)$ are always positive giving rise to monotonic heating. Moreover the bigger is the value of $I(\omega_0)$, the faster is the non-Markovian monotonic heating rate.
For $r=1$, e.g., the monotonic heating occurs at approximately the same rate for all the reservoir types corresponding to the fact that $I(\omega_0)$ is the same for all the reservoirs.

The connection between the low frequencies of the spectrum and oscillations in the heating dynamics can be illustrated by considering the cases where oscillations are present, i.e., for $r\ll 1$. This is the parameter region where $\Delta(t)$ obtains temporarily negative values giving rise to oscillations in $\langle n(t)\rangle$. For these values of $r$ the peaks of all the three spectral distributions are  positioned in the low frequencies region, indicating that the presence of oscillations and the low frequency part of the spectrum are intertwined.

Oscillations in $\langle n(t)\rangle$ mark the presence of non-Markovian memory effects. The persistence of non-Markovian effects for a given value of $r$ (e.g., $r\ll 1$ for high $T$ reservoirs) depends on the type of reservoir spectrum. Figure 6 shows that the memory effects persist for much longer times in the sub-Ohmic reservoir than in the other reservoir types. This indicates that,  when dealing with sub-Ohmic reservoirs, a non-Markovian approach might be needed also at time scales where, for the Ohmic and super-Ohmic environments, a Markovian treatment is sufficient.

We have now illustrated the main features of the non-Markovian heating dynamics of QBM for high temperatures. In the next section we will briefly summarize the corresponding results for the zero temperature case.

\subsubsection{Low Temperatures}

When the system oscillator, initially prepared in its ground state, interacts with a zero temperature
reservoir, the dynamics of the heating
function is basically due to the finite, although small,
system-reservoir coupling energy. At zero temperatures, and for short times, the heating function dynamics is linked to the decay coefficient $\Delta(t)-\gamma(t)$ via Eq. \eqref{eq:appron}.

The non-Markovian dissipative dynamics of QBM at zero $T$ is qualitatively similar to the one shown in Fig. 6 for high $T$ reservoirs.
As discussed in Sec. \ref{decayzeroT}, the coefficient $\Delta(t)-\gamma(t)$, associated to the heating rate of the system oscillator, tends to zero for times much greater than the reservoir correlation time, as expected from the Markovian theory. After the initial non-Markovian heating, the system will eventually thermalize with the zero temperature reservoir.

Similarly to the case of high $T$ reservoirs, also at low temperatures oscillations in $\langle n(t)\rangle$ vanish as $r$ becomes $\gg 1$. Now, however, this typically non-Markovian feature is present for greater values of $r$ than in the high $T$ case, e.g., also for $r=1$. The fact that non-Markovianity is present for a wider range of $r$, in the case of a zero temperature reservoir, is in agreement with what found in Ref. \cite{lindblad} for a Ohmic reservoir with Lorentz-Drude cutoff.

\section{Concluding remarks and future prospects}\label{final}
In this paper we have presented analytical results on the dynamics of a quantum Brownian particle interacting with different types of bosonic thermal reservoirs. Our  approach makes it possible to identify in a clear way the microscopic physical processes taking place at short non-Markovian times scales and to link them to the spectral properties of the environment. By comparing the effects that different types of environments have on the quantum system, we can identify those physical contexts where a non-Markovian approach is required in the description of the time evolution. Moreover we can predict the type of reservoir that perturbs less the quantumness of the system of reference, in our case the harmonic oscillator, and the value of the parameters for which this occurs.  We have seen indeed that the decay and heating rates for the QBM depend strongly on the reservoir type, and in particular on the ratio $r$ between the cutoff frequency $\omega_c$ and the frequency of the system oscillator $\omega_0$.

The time dependent coefficients appearing in the master equation and in the heating function are connected via Eqs. \eqref{eq:appron} and \eqref{eq:highTn} for both zero and high $T$ reservoirs. The heating dynamics for both temperature regimes shows a similar dependence on the parameter $r$. In general oscillations in the heating function, a typical non-Markovian feature, are present for $r\ll 1$. For zero temperatures, however, these oscillations are also present for $r \approx 1$. So for all the types of environment considered, zero $T$ reservoirs are inherently more non-Markovian than high $T$ reservoirs. Moreover, we demonstrated that the sub-Ohmic reservoir induces more pronounced and longer lasting non-Markovian dynamics compared to the other reservoir types and gives rise to a faster Markovian heating in the resonant case ($r=10$).

The oscillations in $\langle n(t)\rangle$ indicate a back and forth exchange of energy between the system and the reservoir. In particular, when the the slope of $\langle n(t)\rangle$ is negative, the system gives back some of the excitation received previously from the reservoir. The statistical  ensemble of system oscillators is, in general, colder than the reservoir with whom it interacts, since we assumed throughout the paper that the system is prepared in its ground state. Oscillations in the heating function therefore indicate the possibility of  a temporary energy flow from a cooler object to a hotter one due to the memory of the environment.

This is not a surprise. The monotonic evolution of the system density operator to its equilibrium value, which is a universal property of quantum dynamical semigroups (Spohn's theorem) \cite{spohn}, is in general violated at short (non-Markovian) timescales. This anomaly has been proposed to be used to control the thermodynamics of an atomic system simply by changing the way in which it is measured \cite{violation_of_thermodynamics}.

Our results indicate that by means of reservoir engineering techniques, e.g., by changing the parameter $r$, one could modify the thermalization dynamics of the system. Another intriguing possibility stemming from reservoir engineering is the simulation of paradigmatic models of open quantum systems as the one discussed in this paper. Schemes for simulating QBM with trapped ions were presented in Refs. \cite{Maniscalco04a,simulator}. A similar approach may be used to simulate the sub-Ohmic and super-Ohmic environments here investigated.

An ideal physical context where our results could be experimentally verified is the trapped ion context.
Experiments with single trapped ions have demonstrated the ability to engineer artificial environments and to control the relevant system-environment parameters \cite{engineeredres}. These experiments aim at measuring the decoherence of a quantum superposition of coherent states and Fock states due to
the presence of the reservoir. Several types of engineered reservoirs are demonstrated, e.g., thermal amplitude reservoirs, phase reservoirs, and zero temperature reservoirs \cite{engineeredres}.

A high $T$ amplitude reservoir is obtained by applying a random
electric field $\vec{E}$ whose spectrum is centered on the axial
frequency $\omega_z $  of oscillation of the ion. The trapped ion motion couples to this field
due to the net charge $q$ of the ion: $H_{int}= -q \vec{x} \cdot
\vec{E}$, with $\vec{x}=(X,Y,Z)$ displacement of the  c.m. of the
ion from its equilibrium position. Remembering that $\vec{E}
\propto \sum_i \vec{\epsilon}_i (b_i + b^{\dag}_i)$, with $b_i$
and $b^{\dag}_i$ annihilation and creation operators of the
fluctuating field modes, and that $X\propto \left( a +a^{\dag}
\right)$ the quantized position operator of the ion motion, one realizes that this coupling is equivalent to the
bilinear one given in Eq. (\ref{eq:interactionhamilton}).

The random electric field is applied to the endcap electrodes
through a network of properly arranged low pass filters limiting
the \lq\lq natural\rq\rq~environmental noise but allowing
deliberately large applied fields to be effective. This type of
drive simulates an infinite-bandwidth amplitude reservoir
\cite{engineeredres}. It is worth stressing that, for the times of
duration of the experiment the heating due to the natural reservoir is definitively negligible
\cite{engineeredres}.

The different high $T$ spectra that we
discuss in this paper can be realized experimentally by filtering
the random field, used in the experiments for simulating an
infinite-bandwidth reservoir, with a suitable set of band-pass filters.
This enables the comparison of the heating rates between different reservoir spectra and the observation of non-Markovian effects. We notice that measurements of the heating function are routinely performed in the trapped ion context. The heating function is obtained from measurements of the population of the vibrational states of the ion. A detailed study of the experimental techniques to simulate harmonic quantum Brownian motion with trapped ions, for the case of an Ohmic spectrum with Lorentz-Drude cutoff has been performed in Ref. \cite{Maniscalco04a}. We believe the methods needed to verify these phenomena are already in the grasp of the experimentalists \cite{Maniscalco04a,simulator}.


Finally, it is worth recalling that
manipulation of the dynamics via certain types of measurements causes the decay processes to be inhibited or accelerated, depending on the system-reservoir properties. These crucially quantum phenomena, known as quantum Zeno (QZE) and anti-Zeno effects (AZE), have been studied for the QBM model in the case of an Ohmic spectrum in Ref.  \cite{zeno}.

The results presented in this paper pave the way to the study of the influence of the reservoir spectrum on the occurrence of quantum
Zeno or anti-Zeno effect (AZE).
The borderline between the occurrence of QZE or AZE is indeed related to the spectral properties of the environment and therefore will depend on the type of reservoir (Ohmic, sub-Ohmic, super Ohmic) considered. A comparative study of the
Zeno-anti-Zeno crossover is thus a natural follow up of this paper.

\acknowledgments

This work has been  supported by the Emil Aaltonen foundation, the Vilho, Yrj\"o and Kalle V\"ais\"al\"a foundation, the Magnus Ehrnrooth
Foundation, and the Academy of
Finland (projects 108699, 115982, 115682, 8125004). S.M. also thanks the Turku Collegium of Science and Medicine for financial support.

\end{document}